\documentclass[letter]{aa} 
\usepackage{threeparttable}
\usepackage{graphicx}
\usepackage{amsmath}
\usepackage{txfonts}
\usepackage{natbib}
\usepackage{braket}
\usepackage{tabularx}

\newcommand{\eg}{e.g.,\ }

\begin{document} 
\makeatletter
\makeatother

\title{Heavy element enrichment of gas in surface-accretion disks} 
\subtitle{A possible origin of the mass--metallicity anti-correlation in exoplanets}

\author{Yoshitaka Ikeda
          \inst{1}
          \and
          Kazumasa Ohno
          \inst{2}
          \and
          Satoshi Okuzumi
          \inst{1}
          }

   \institute{
   Department of Earth and Planetary Sciences, Tokyo Institute of Technology, Meguro, Tokyo, 152-8551, Japan
        \and
    National Astronomical Observatory of Japan, Osawa, Mitaka, Tokyo, 181-8588, Japan\\
             \email{ikeda.y.75f5@m.isct.ac.jp}
             }

\date{}
\abstract{
   Recent observations, including those by JWST, suggest that the atmospheres of many gas giant exoplanets have super-stellar metallicity that is anti-correlated with planetary mass. 
   Several studies suggest that the super-stellar metallicity can be explained by accretion of vapor-enriched disk gas produced by the sublimation of rapidly drifting icy pebbles.
   However, recent disk observations and experiments suggest that icy dust is fragile at low temperatures, calling into question the conventional picture that icy grains grow efficiently and drift rapidly.
   }
   {
    We present a new scenario for heavy-element enrichment in the inner disk by fragile, slowly drifting icy dust, assuming that magnetohydrodynamical disk winds drive gas accretion near the disk surface rather than at the midplane.
   }
   {
   We simulate the evolution of gas and dust in a surface-accretion disk, taking into account the radial transport of gas and dust, collision growth and fragmentation of fragile dust, and the condensation and sublimation of H$_2$O.
   Two accretion disk models are presented, in which gas accretion flows are assumed to be either vertically uniform or narrowly concentrated near the disk surface.
   }
  {
  In the uniform accretion disk model, fragile icy grains enhance the water vapor abundance inside the snow line only by a factor of ${\sim}3$ due to their slow drift.
  In contrast, in the surface-accretion disk model, the slow drift of icy dust leads to water vapor enrichment that is higher by an order of magnitude, owing to the selective removal of ice-free gas from the disk. 
  Furthermore, surface accretion yields an anti-correlation between the water vapor concentration in the inner disk and the residual disk gas mass, analogous to the anti-correlation between atmospheric metallicity and planet mass observed in extrasolar giant planets.
  }
  {
  Surface gas accretion naturally establishes vapor-rich environments inside the snow line when icy dust is fragile.
  This study provides a novel perspective on the formation environments that dictate the composition of gas giant atmospheres.
  }


\maketitle
\nolinenumbers
\section{Introduction}
\begin{figure}[t]
\centering
\includegraphics[width=0.95\columnwidth,bb=0 0 576 432]{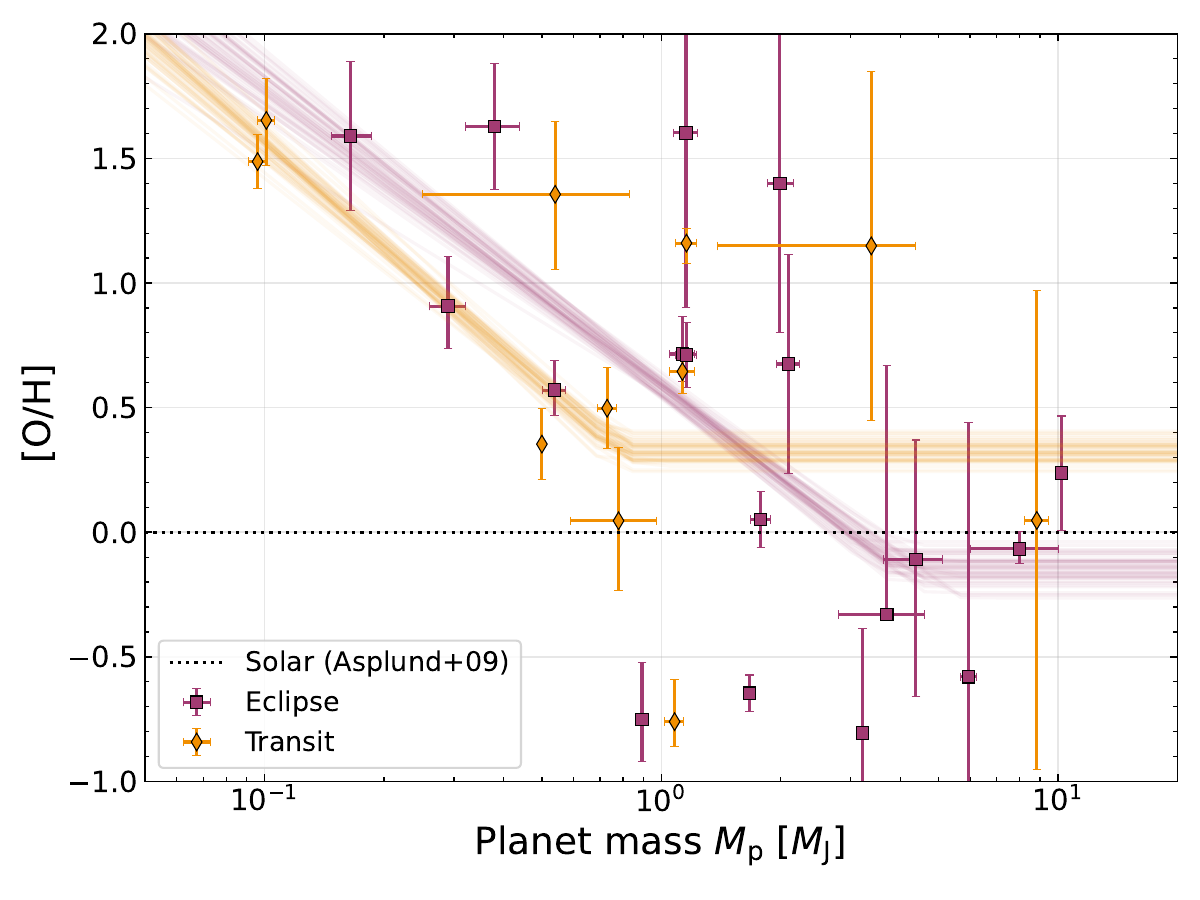}
\caption{
Atmospheric oxygen abundances for hot to warm giant exoplanets as a function of planet mass (in units of Jupiter mass $M_{\rm J}$). 
The abundances are expressed as $[\rm O/H]=\log_{10}(O/H)-\log_{10}(O/H)_{\odot}$, adopting solar abundances from \citet{Asplund2009}.
Purple and orange points denote values inferred from emission and transmission observations. 
The data and fitted curves are compiled from the ExoComp database \citep{Lothringer2026}.
}

\label{fig:intro}
\end{figure}

Atmospheric elemental abundances of gas giant planets provide an important constraint on our understanding of gas giant planet formation.
Observations of exoplanet atmospheres, including recent results from JWST, reveal that many gas giant exoplanets have super-stellar atmospheric metallicities, and some of them have the metallicities that are more than an order of magnitude higher than their host stars \citep[\eg][]{Feinstein2023,Bean2023,Fu2024}.
Recent studies also suggest that atmospheric metallicity is anti-correlated with planetary mass \citep[\eg][]{Welbanks19,Kempton24,Fu2025,Lothringer2026}, as illustrated in Fig.~\ref{fig:intro}.

One possible origin of such high bulk/atmospheric metallicities is the enrichment of the disk gas with heavy elements through the sublimation of the icy dust inside the snow line. 
Sublimation of radially drifting dust grains enriches the local disk gas in heavy elements. 
If a gas giant accretes this enriched gas during its formation, its atmosphere will inherit the metallicity enhancement \citep[\eg][]{Booth2017,Schneider21,Ohno2026}.
Alternatively, accretion of planetesimals to the primitive planetary atmosphere can enrich the planetary metallicities \citep[\eg][]{Fortney2013,Shibata2023}.
Therefore, atmospheric metallicity provides an important clue to the evolution of the gas composition as well as to the spatial and size distributions of planetesimals in the planet-forming disks.

Previous models of disk gas enrichment assumed that icy dust has a high sticking efficiency \citep[e.g.,][]{DominikTielens97}, enabling efficient growth and rapid radial drift that enrich the inner disk with water vapor. 
However, recent disk observations \citep{OkuzumiTazaki19,Jiang24,Ueda24} and experiments \citep{Gundlach2018,Musiolik2019} suggest that icy dust at low temperatures might be less sticky, and therefore drift more slowly, than previously assumed. 
These studies call into question the previous idea that efficient dust growth and rapid inward drift enrich the disk gas.

Recently, \citet{Okuzumi2025} proposed a scenario in which slow dust drift  can still lead to dust enrichment in a gas disk.  
The proposed scenario assumes that the disk's gas accretion occurs on its surface rather than the conventionally assumed midplane.
Such surface accretion has been observed in recent magnetohydrodynamic simulations with realistic ionization distribution \citep[e.g.,][]{Gressel15,Bai17,Lesur21,Iwasaki24}.
Gas accretion concentrated near the disk surface is due to enhanced ionization and magnetic coupling in the surface layers.
If dust is settled below this surface layer, the accretion flow removes gas, but not dust, from the disk.
Assuming that dust grains are fragile and drift slowly, \citet{Okuzumi2025} demonstrated that the selective removal of gas by surface accretion results in an increase of the disk's dust-to-gas ratio with decreasing disk gas mass.

In this Letter, we show that the selective removal of gas due to surface accretion, combined with ice sublimation from slowly drifting grains crossing the snow line, also leads to metallicity enrichment of the inner disk {\it gas}. This occurs because the enhanced solid-to-gas mass ratio outside the snow line increases the ice-to-gas flux ratio across it. We demonstrate that this mechanism produces an anti-correlation between the disk gas mass and the metallicity of the inner disk region,  which may be related to the mass--metallicity anti-correlation of close-in giant exoplanets.

\begin{figure}[t]
\centering
\includegraphics[width=1.0\hsize, bb= 0 0 723 872]{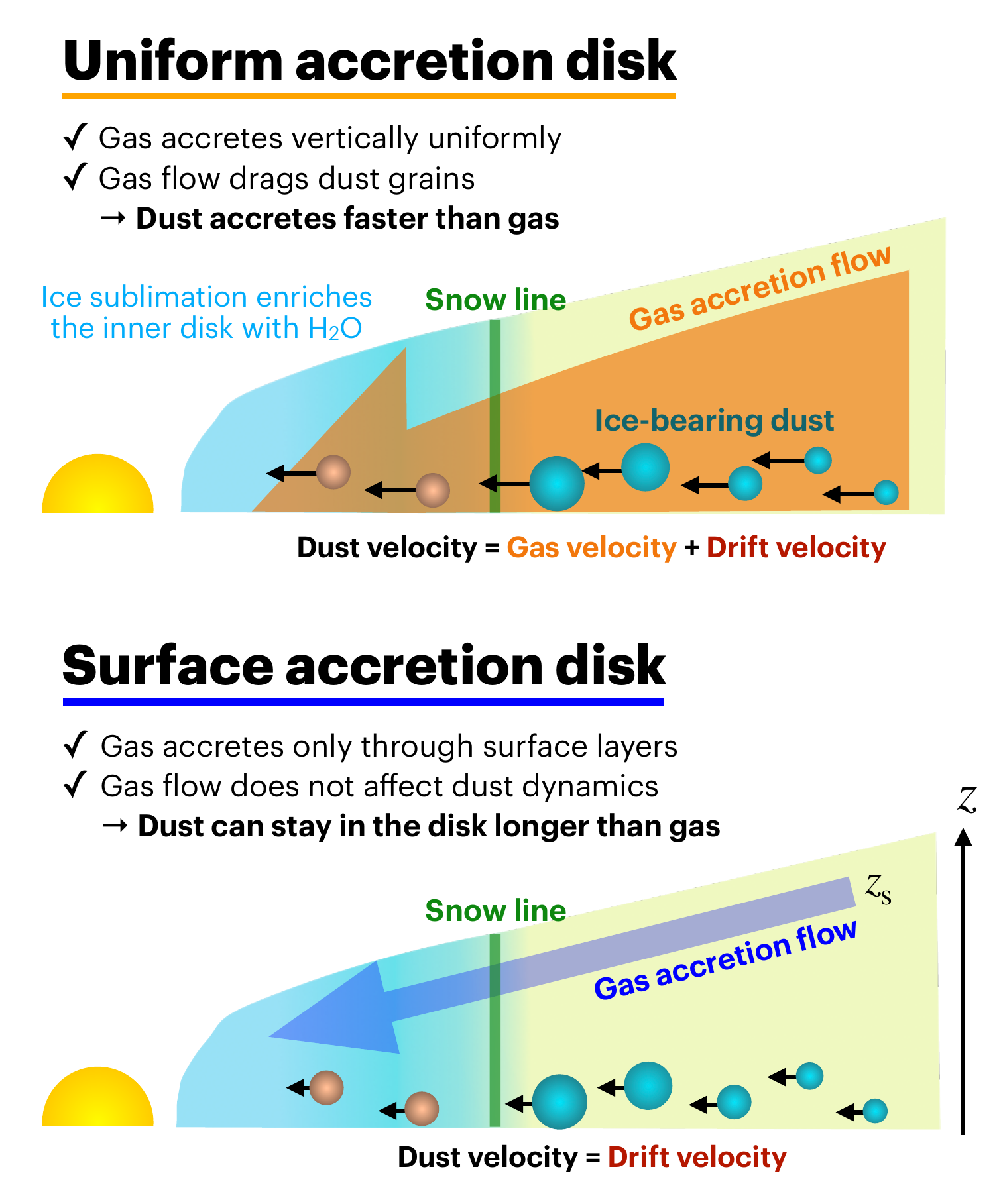}
\caption{
Schematic illustration of gas and dust transport in disks with vertically uniform and surface accretion (upper and lower panels, respectively).
Vertically uniform gas transports dust at all heights, whereas accretion concentrated at height (denoted by $z =z_{\rm s}$) does not transport dust below this height.
As ice-bearing dust grains in the outer disk region move in and cross the snow line (vertical line), they release water vapor and thereby enhance the inner disk's metallicity.
See Appendix \ref{Appendix_model} for details of the adopted disk models.
}
\label{fig:model}
\end{figure}

\begin{figure*}[t]
\centering
\includegraphics[width=0.95\hsize,bb=0 0 1920 780]{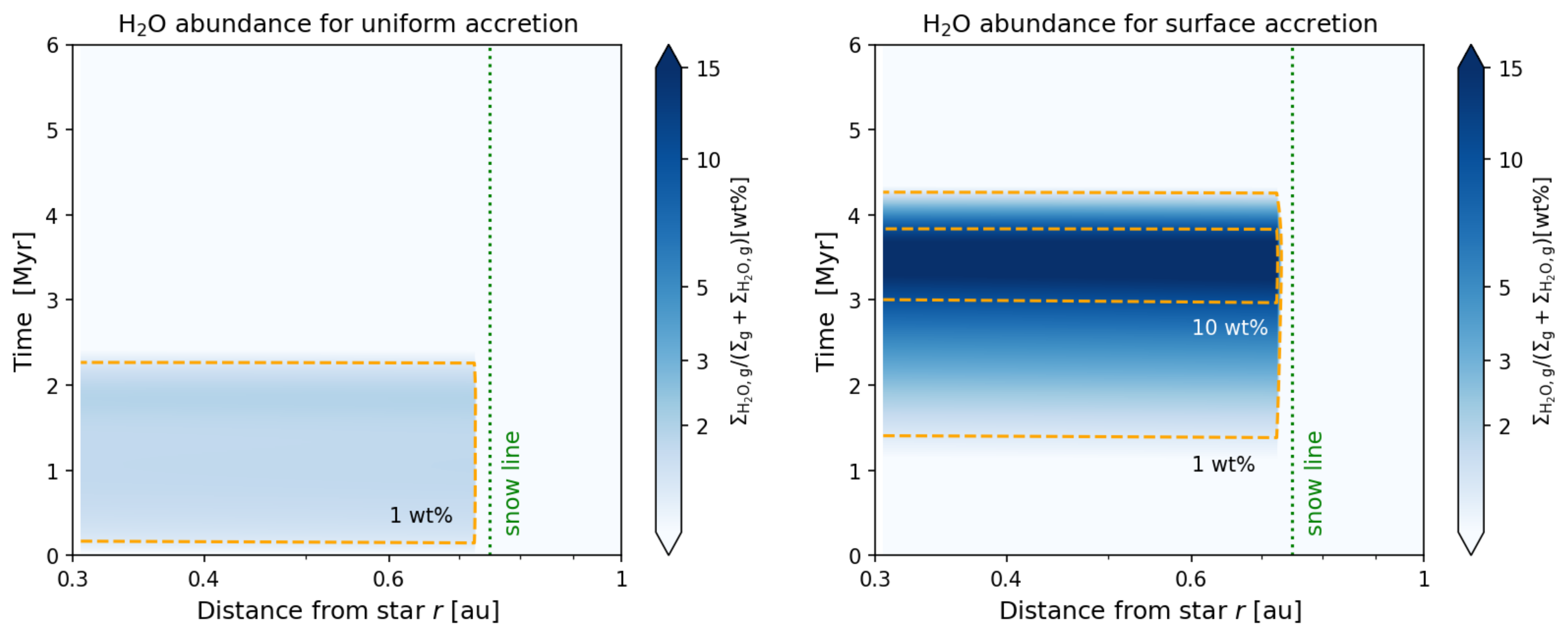}
\caption{
Evolution of the water vapor concentration $\Sigma_{\rm H_2O,g}/(\Sigma_{\rm g}+\Sigma_{\rm H_2O,g})$ in the uniform and surface-accretion disk models (left and right panels, respectively) with $v_{\rm stick} = 0.3~\rm m~s^{-1}$.
The dashed lines indicate vapor concentrations of 1 and 10~wt\%.
The dotted line marks the snow line.}
\label{fig:enrichment}
\end{figure*}

\section{Method}\label{sec:Method}
We investigate the H$_2$O vapor concentration inside the snow line by simulating the time evolution of gas, dust, and H$_2$O surface densities, as detailed in Appendix \ref{Appendix_model}.
We assume that disk gas accretion is mainly mediated by magnetically driven winds \citep{Tabone22}.
We consider two disk models in which the wind-driven accretion flow is either vertically uniform or concentrated near the disk surface (referred to as the uniform and surface-accretion disks, respectively; see Fig.~\ref{fig:model}). 
In both models, a low level of vertically uniform viscosity-driven accretion is also included, but wind-driven accretion dominates the net accretion rate. Turbulent dust diffusion is also included.

In general, the vertical profile of gas accretion flows has a critical impact on heavy-element fractionation in disks, because gas and dust have different vertical distributions.
In our models, vertically uniform accretion flows transport dust, but surface accretion flows do not, assuming that the dust is settled below the surface layer (for more discussions, see \citealt{Okuzumi2025}).
In the uniform accretion disk (the upper panel of Fig.~\ref{fig:model}),
the effective radial transport velocity for the dust surface density, defined as the dust-density-weighted vertical average of the radial velocity, can be expressed as (Eq. (36) of \citealt{Okuzumi2025})
\begin{equation}
  \langle v_{{\rm d},r} \rangle_{\rm d,Uni}\approx \frac{ \langle v^{\rm visc}_{{\rm g},r}\rangle_{\rm g} +\langle v^{\rm wind}_{{\rm g},r}\rangle_{\rm g}}{1+\rm St_{\rm mid}^2}  + \frac{2{\rm St_{\rm mid}}\Delta v_{\rm g,\phi,mid}}{1+\rm St_{\rm mid}^2}.
\label{eq:vdr_uni}
\end{equation}
Here, $\langle v^{\rm visc}_{{\rm g},r}\rangle_{\rm g} $ and $ \langle v^{\rm wind}_{{\rm g},r}\rangle_{\rm g}$ are the effective radial transport velocity by viscosity and disk wind, $\rm St_{\rm mid}$ is the Stokes number of dust in the midplane,
and $\Delta v_{\rm g,\phi,mid}$ is the rotation velocity of the gas relative to Keplerian motion. 
In the surface-accretion disk (the lower panel of Fig.~\ref{fig:model}), 
the corresponding transport velocity is (Eq.~(37) of \citealt{Okuzumi2025})
\begin{equation}
    \langle v_{{\rm d},r} \rangle_{\rm d,Sur}\approx \frac{\langle v^{\rm visc}_{{\rm g},r}\rangle _{\rm g}}{1+\rm St_{\rm mid}^2}  + \frac{2{\rm St_{\rm mid}}\Delta v_{\rm g,\phi,mid}}{1+\rm St_{\rm mid}^2}.
\label{eq:vdr_sur}
\end{equation}
In Eqs.~\eqref{eq:vdr_uni} and \eqref{eq:vdr_sur}, the first term represents co-accretion with the vertically uniform flows, whereas the second term represents the radial drift relative to the gas.
Unlike in the uniform accretion model, wind-driven accretion does not contribute to radial dust transport in the surface accretion model. This is the key difference between the two models.
The assumption that dust is well settled below the surface accretion layer is valid in the parameter space studied in this study (see Appendix~\ref{Appendix:gasdust} and \ref{appendix:result}).

We evolve the gas and dust surface densities by solving  radially one-dimensional advection--diffusion equations (Eqs.~\eqref{eq:gas_surfacedensity_v} and \eqref{eq:dust_surfacedensity}).
Following \citet{Tabone22}, we express $\langle v^{\rm visc}_{{\rm g},r}\rangle_{\rm g}$ and $\langle v^{\rm wind}_{{\rm g},r}\rangle_{\rm g}$ by dimensionless parameters, $\alpha_{\rm visc}$ and $\alpha_{\rm wind}$ (Eqs.~\eqref{eq:vgr_visc} and \eqref{eq:vgr_wind}).
We set $\alpha_{\rm visc}=3\times{10}^{-4}$ and $\alpha_{\rm wind}=6\times{10}^{-3}$ throughout this paper to enforce the same evolution of the gas surface density in the uniform and surface accretion models. The initial disk mass is assumed to be $0.1M_\odot$.

Our model also accounts for the collisional growth and fragmentation of dust grains. The sticking threshold $v_{\rm stick}$ defines the collision velocity below which grains grow in mass; at higher collision velocities, grains are assumed to fragment and lose their individual mass while conserving the total dust mass (see Eq.~\eqref{eq:appendix_xi}).
We adopt $v_{\rm stick} = 0.3~{\rm m~s^{-1}}$ as the fiducial value and also consider $v_{\rm stick} = 1~{\rm m~s^{-1}}$ to examine the impact of dust fragility. 
These choices are motivated by the low values of $v_{\rm stick} \lesssim 1~{\rm m~s^{-1}}$ inferred from multi-wavelength modeling of several disks \citep{OkuzumiTazaki19,Jiang24,Ueda24}.

To treat heavy-element enrichment in the disk, we also evolve the total surface density $\Sigma_{\rm H_2O}$ of H$_2$O (i.e., the sum of ice and vapor).
The advection--diffusion equation for $\Sigma_{\rm H_2O}$ (Eq.~\eqref{eq:h2o_surfacedensity}) assumes small dust and instantaneous ice sublimation and condensation across the snow line (see Appendix \ref{Appendix_model} for details).
Inside and outside the H$_2$O snow line, the advection velocity and diffusion coefficient for H$_2$O are taken to be those for gas and dust, respectively.
The snow line is fixed to the orbit at which $T = 160~\rm K$.

\section{Results}\label{sec:Results}
Fig.~\ref{fig:enrichment} shows the water vapor concentration, defined as the ratio of the water vapor surface density to the total gas surface density, $\Sigma_{\rm H_2O,g}/(\Sigma_{\rm g}+\Sigma_{\rm H_2O,g})$, in two models. 
In the case of the uniform accretion disk (the left panel of Fig.~\ref{fig:enrichment}), the vapor concentration inside the snow line reaches $2~\rm wt\%$ at 2~Myr.
This value is four times higher than the initial value of 0.5 wt\ but is considerably lower than the maximum enrichment of $\sim10~\rm wt\%$ predicted by previous models \citep[see e.g., Fig.~4 of][]{Schneider21}.
This is because our model assumes  more fragile dust ($v_{\rm stick} =0.3~\rm m~s^{-1}$), resulting in smaller fragmentation-limited sizes (corresponding to ${\rm St} \sim 10^{-4}$--$10^{-3}$; see Fig.~\ref{fig:surface_density}) and hence slower inward drift.
As a result, our uniform accretion model predicts lower inward H$_2$O flux across the snow line than the previous studies. 
This result demonstrates that fragile icy dust, as suggested by recent observations and experiments, is inefficient at enriching the inner disk with water vapor, as long as accretion is vertically uniform.

In contrast, in the surface-accretion disk (Fig.~\ref{fig:enrichment}, right), the water vapor concentration inside the snow line increases more significantly, reaching 15~wt\% at 3.5~Myr.
Since surface accretion does not transport dust, the surface density of the slowly drifting, ice-bearing dust decreases more slowly than that of the gas (see Fig.~\ref{fig:surface_density} in Appendix \ref{Appendix:abundance}).  
The retention of ice-bearing dust leads to a relative enhancement of the ice-to-gas mass ratio outside the snow line, increasing the water-to-gas flux ratio across the snow line and thereby enhancing the water vapor abundance interior to it.
Appendix \ref{Appendix:abundance} provides an analytical argument to help understand the present results. The water vapor concentration starts to decrease only at $t \sim 4~\rm Myr$, which corresponds to the timescale of dust loss due to inward drift in this model (see Sect.~5.1 of \citealt{Okuzumi2025}).

\begin{figure}[t]
\includegraphics[width=\columnwidth, bb=0 0 360 288]{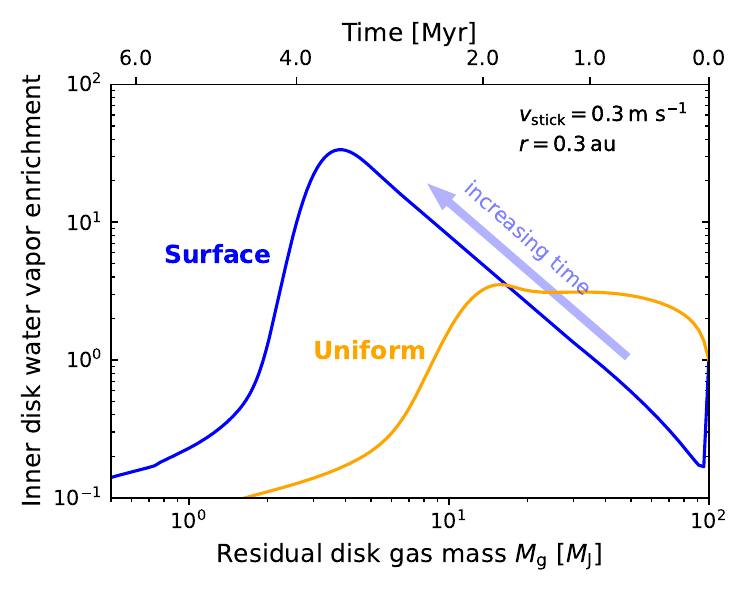}
\vskip-10pt
\includegraphics[width=\columnwidth,bb=0 0 360 288]{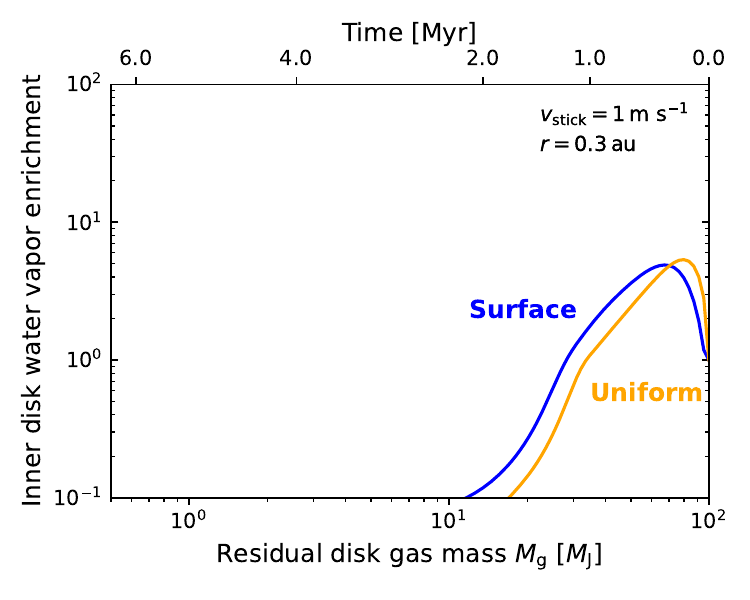}
\vspace{-8mm}
\caption{
Water vapor enrichment (see text for definition) in the inner disk, measured at $r=$ 0.3~au, as a function of the residual gas mass of the entire disk, $M_{\rm g}$. 
The time corresponding to the residual disk gas mass is shown on the top horizontal axis.
The orange and blue lines are for the uniform and surface-accretion disk models.
The upper and lower panels compare these models for $v_{\rm stick}=0.3$ and $1~\rm m~s^{-1}$.
}
\label{fig:enrichment_to_initial}
\end{figure}

Since the disk gas mass decreases monotonically with time, the water vapor enrichment inside the snow line can be plotted as a function of residual disk gas mass, $M_{\rm g}(t)$.
Fig.~\ref{fig:enrichment_to_initial} displays the water vapor enrichment at $0.3~\mathrm{au}$ versus $M_{\rm g}$ for different disk models.
Here, we define the water vapor enrichment as the water vapor concentration $\Sigma_{\rm H_2O,g}/(\Sigma_{\rm g}+\Sigma_{\rm H_2O,g})$ normalized by its initial value of 0.5 wt\%. 
Assuming $v_{\rm stick}=0.3~{\rm m~s^{-1}}$ (upper panel), the water vapor enrichment in the uniform accretion disk model plateaus out at a factor of a few as soon as $M_{\rm g}$ decreases from the initial value of $\approx 100 M_{\rm J}$. This plateau value is set by the ratio between the dust and gas radial velocities in the outer disk, at which the inward dust and gas flows originate. 
In contrast, in the surface-accretion disk model, water vapor enrichment continues to increase as $M_{\rm g}$ decreases, reflecting the fact that $\Sigma_{\rm H_2O,g}$ declines more slowly than $\Sigma_{\rm g}$. This anti-correlation between vapor enrichment and $M_{\rm g}$ persists until $M_{\rm g}$ falls below $\sim 3M_{\rm J}$, corresponding to $t \sim 4~\rm Myr$, after which the ice-transporing dust is depleted by radial drift.

The water vapor enrichment discussed above depends on the assumed dust fragility.
The lower panel of Fig.~\ref{fig:enrichment_to_initial} shows the water vapor enrichment in models with stickier grains of $v_{\rm stick}=1~{\rm m~s^{-1}}$) as adopted in previous studies.
These models exhibit similar evolutionary trends in water vapor enrichment,  because the stickier grains grow larger and their inward motion is dominated by radial drift (the last term in Eqs.~\eqref{eq:vdr_uni} and \eqref{eq:vdr_sur}) rather than co-accretion with gas.
Notably, as $v_{\rm stick}$ is decreased, the peak enrichment value in the uniform accretion model decreases, whereas that in the surface accretion model increases.
Thus, while the conventional uniform accretion disk model requires stickier icy dust to explain the metallicity in the disk gas, our surface accretion model attains metallicity enrichment when the grains are fragile.

\section{Conclusions}\label{sec:Discussion}
In this Letter, we have presented, for the first time, gas--dust evolution models that account for both surface disk accretion and ice sublimation across the snow line.
We found that the inner region of a surface-accretion disk could be significantly enriched in H$_2$O vapors if the ice-bearing dust is fragile, as suggested by recent disk observations.
The gas metallicity also becomes anti-correlated with the residual disk gas mass (upper panel of Fig.~\ref{fig:enrichment_to_initial}).

Intriguingly, this anti-correlation is qualitatively similar to the mass–metallicity anti-correlation inferred for gas giants \citep[\eg][see also Fig.~\ref{fig:intro}]{Welbanks19,Kempton24,Fu2025,Lothringer2026}. Specifically, under the simple assumption that the final mass of a gas giant scales with the residual disk gas mass, the surface-accretion model shown in the upper panel of Fig.~\ref{fig:enrichment_to_initial} reproduces the observationally inferred trend in which the atmospheric metallicity of giant planets increases by up to a factor of several tens as the planet mass decreases by a comparable factor. Such a strong anti-correlation is not readily expected in uniform-accretion disk models, where the metallicity in the disk gas tends to either remain constant or decline as the residual disk mass decreases (see Fig.~\ref{fig:enrichment_to_initial}).
Indeed, recent planet formation simulations by \citet{Ohno2026} demonstrate that planets accreting vapor-enriched disk gas inside the water snow line exhibit only a weak mass–metallicity anti-correlation. Future planet formation models that account for surface accretion will provide a direct test of our proposed scenario, in which gas giants with anti-correlated atmospheric mass and metallicity form in surface-accretion disks.

In addition, JWST observations of inner-disk molecular lines \citep[e.g.,][]{Long2025,Banzatti2025,Arulanantham2025} have begun to reveal trends in inner-disk chemistry, including signatures of inward pebble drift. These observations may serve as important testbeds for our model. However, direct comparison with the model requires careful consideration of the observed disks' radial substructures (i.e., gaps and rings), which may partially trap drifting pebbles. The disks' vertical structure is also crucial, as infrared observations primarily trace the disk surface rather than the midplane. Future modeling efforts addressing these issues will be important for testing the enrichment process proposed in this study.

\begin{acknowledgements}
We thank Shota Notsu for helpful discussions.
This work is supported by JSPS KAKENHI Grant number JP23K25923, JP23K19072, and JP26K17222.
Figure 1 made use of the ExoComp database \citep{Lothringer2026}.
\end{acknowledgements}

\bibliographystyle{aa}

\appendix
{ 
\section{Model details}\label{Appendix_model}
In this Appendix, we describe our models for the evolution of gas and dust in uniform and surface-accretion disks, as well as the condensation and sublimation model.

\subsection{Gas and dust transport}
\label{Appendix:gasdust}
Following \citet{Tabone22} and \citet{Okuzumi2025}, we express the time evolution of the gas surface density as

\begin{equation}
    \frac{\partial \Sigma_{\rm g}}{\partial t} = -\frac{1}{r}\frac{\partial}{\partial r}\bigg[ r(\langle v^{\rm visc}_{{\rm g},r}\rangle_{\rm g} +\langle v^{\rm wind}_{{\rm g},r}\rangle_{\rm g})\Sigma_{\rm g} \bigg] -\dot{\Sigma}^{\rm wind}_{\rm g},
    \label{eq:gas_surfacedensity_v}
\end{equation}
\begin{equation}
    \dot{\Sigma}^{\rm wind}_{\rm g}=\frac{3\alpha_{\rm wind}c_s^2 \Sigma_{\rm g}}{4(\lambda -1)r^2 \Omega}\label{eq:Sigma_dot}.
\end{equation}
Where $\Omega$ is the angular velocity, $c_s$ is the isothermal sound speed, and $\lambda$ is the magnetic arm parameter.
The sound speed is determined from the disk temperature profile, for which we assume stellar irradiation to be the dominant heating source, following equation (18) of \citet{Okuzumi2025}.
For simplicity, we calculate the sound speed $c_s$ is assuming a fixed mean molecular weight of $\mu =2.3~m_{\rm H}$, where $\mu$ and $m_{\rm H}$ denote the mean molecular weight and hydrogen mass, respectively.
$\Sigma_{\rm g}$ represents the surface density of disk excluding $\rm H_2O$ vapor.

The effective gas accretion velocities due to turbulent viscosity and disk wind, $\langle v^{\rm visc}_{{\rm g},r}\rangle_{\rm g}$ and $\langle v^{\rm wind}_{{\rm g},r}\rangle_{\rm g}$, are given by \citep[equation (22) and (23) in][]{Okuzumi2025}

\begin{equation}
    \langle v^{\rm visc}_{{\rm g},r}\rangle_{\rm g}=-\frac{3\alpha_{\rm visc} c^2_s}{r\Omega}\frac{\partial \ln (r^2\alpha_{\rm visc}\Sigma_{\rm g} c_s^2) }{\partial \ln r},
\label{eq:vgr_visc}
\end{equation}
\begin{equation}
    \langle v^{\rm wind}_{{\rm g},r}\rangle_{\rm g}=-\frac{3\alpha_{\rm wind} c^2_s}{2 r \Omega}.
\label{eq:vgr_wind}
\end{equation}
Here, $\alpha_{\rm visc}$ represents the Shakura--Sunyaev viscosity parameter \citep{Shakura73}, while $\alpha_{\rm wind}$ is a dimensionless quantity characterizing the strength of the magnetohydrodynamic (MHD) wind torque responsible for the removal of the angular momentum from the disk \citep[equivalent to $\alpha_{\rm DW}$ in][]{Tabone22}.
We assume $\alpha_{\rm visc}=3 \times10^{-4}$ and $\alpha_{\rm wind}=6\times 10^{-3} $ throughout in this paper.
The total vertically averaged gas velocity $\langle v_{{\rm g},r}\rangle_{\rm g}$ can be expressed as the sum of the individual contributions,$\langle v_{{\rm g},r}\rangle_{\rm g}=\langle v^{\rm visc}_{{\rm g},r}\rangle_{\rm g} +\langle v^{\rm wind}_{{\rm g},r}\rangle_{\rm g}$.
The effective radial velocity is defined as the density-weighted vertical average of the local radial velocity and is given by
\begin{equation*}
    \langle X \rangle_{\rm {g} }(r) \equiv \frac{1}{\Sigma_{\rm g}(r)}\int X(r,z)\rho_{\rm g}(r,z)dz,
\end{equation*}  
where $X=v^{\rm visc}_{{\rm g},r}$ or $v^{\rm wind}_{{\rm g},r}$, and $\rho_{\rm g}$ is the gas density.
We note that the averaged velocity is associated with the gas accretion rate through $\dot{M}_{\rm g}= 2\pi r\Sigma_{\rm g}\langle v_{{\rm g},r}\rangle_{\rm g}$.

Following \citet{Okuzumi2025}, we express the time evolution of the dust surface density 
$\Sigma_{\rm d}$ as
\begin{equation}
    \frac{\partial \Sigma_{\rm d}}{\partial t} = -\frac{1}{r}\frac{\partial}{\partial r} r \bigg[\langle v_{{\rm d},r} \rangle_{\rm d} \Sigma_{\rm d} - \Sigma_{\rm g} D_{{\rm d},r}\frac{\partial}{\partial r} \bigg(\frac{\Sigma_{\rm d}}{\Sigma_{\rm g}}\bigg ) \bigg ]\label{eq:dust_surfacedensity},
\end{equation}
where $\langle v_{{\rm d},r} \rangle_{\rm d}$ and $D_{{\rm d},r}$ are the effective radial velocity and radial diffusion coefficient for dust, respectively. In our definition, $\Sigma_{\rm d}$ only accounts for refractory components, excluding $\rm H_2O$.

The dust velocity and diffusion coefficient depend on the Stokes number of dust grains, $\rm St$. Assuming Epstein's drag, the midplane Stokes number ${\rm St}_{\rm mid}$ scales with the grain radius $a$ as
\begin{equation}
    {\rm St_{mid}} =\frac{\pi}{2}\frac{\rho_{\rm int} a}{\Sigma_{\rm g}},\label{eq:St}
\end{equation}
where $\rho_{\rm int}$ is the internal density of the dust particles, which we fix to be $0.6~\rm g~cm^{-3}$.
The effective dust radial velocity $\langle v_{{\rm d},r} \rangle_{\rm d}$ is given by 
\begin{equation}
    \langle v_{{\rm d},r} \rangle_{\rm d} = \left\langle \frac{v_{{\rm g},r}}{1+\rm St^2} \right\rangle_{\rm d} + \frac{2{\rm St_{\rm mid}}\Delta v_{\rm g,\phi,mid}}{1+\rm St_{\rm mid}^2},
    \label{eq:vdr_base}
\end{equation}
where the brackets $\langle \cdots \rangle_{\rm d}$ represents a vertical average weighted by the vertical dust distribution (analogous to $\langle \cdots \rangle_{\rm g}$) and
$\Delta v_{{\rm g},\phi,{\rm mid}}
= (c_s^2 / 2 r \Omega)
(\partial \ln P / \partial \ln r)$ denotes the deviation of the midplane azimuthal velocity in disk gas from the Keplerian velocity.
For simplicity, we assume that the gas velocity induced by viscosity is vertically uniform, i.e., $v^{\rm visc}_{{\rm g},r}(z)=\langle v^{\rm visc}_{{\rm g},r}\rangle_{\rm g}$. The dust diffusion coefficient is given by $D_{{\rm d},r} = D_{{\rm g},r}/(1+{\rm St^2_{mid}})$, where $D_{{\rm g},r} = \alpha_r c_{\rm s}^2/\Omega$ is the gas diffusion coefficient and $\alpha_r$ is the corresponding dimensionless diffusion coefficient.

The main difference between uniform and surface-accretion disks is the vertical distribution of the wind-driven gas flow.
In uniform accretion disks, we assume that the wind-driven gas accretion flow is vertically uniform as assumed for viscous accretion flow, i.e., $v_{{\rm g},r}^{\rm wind}=\langle v^{\rm wind}_{{\rm g},r}\rangle_{\rm g}$. 
Thus, the radial dust velocity is given by
\begin{equation}
    \langle v_{{\rm d},r} \rangle_{\rm d,Uni}\approx \frac{ \langle v^{\rm visc}_{{\rm g},r}\rangle_{\rm g} +\langle v^{\rm wind}_{{\rm g},r}\rangle_{\rm g}}{1+\rm St_{\rm mid}^2}  + \frac{2{\rm St_{\rm mid}}\Delta v_{\rm g,\phi,mid}}{1+\rm St_{\rm mid}^2},
    \label{eq:vdr_uni_appendix}
\end{equation}
In contrast, in the surface-accretion disk, we assume that the gas flow is localized at height $z=z_s$ (the lower panel of Fig.~\ref{fig:model}).
Such localized gas flow is suggested by magnetohydrodynamic (MHD) simulations \citep{Fu&Bai2021}.
Therefore, the effective radial velocity of dust, weighted by the dust mass density, is given by \citep[see Sec.~3.3.2 of][]{Okuzumi2025}
\begin{equation}
    \langle v_{{\rm d},r} \rangle_{\rm d,Sur}\approx C_{\rm surface}\frac{\langle v^{\rm wind}_{{\rm g},r}\rangle _{\rm g}}{1+{\rm St}^2(z_{\rm s})}+\frac{\langle v^{\rm visc}_{{\rm g},r}\rangle _{\rm g}}{1+\rm St_{\rm mid}^2}  + \frac{2{\rm St_{\rm mid}}\Delta v_{\rm g,\phi,mid}}{1+\rm St_{\rm mid}^2}.
    \label{eq:vdr_Csurface}
\end{equation}
Here, $C_{\rm surface}$ is the dust-to-gas ratio at $z = z_{\rm s}$ normalized by the dust-to-gas surface density ratio.
Assuming the balance between the vertical settling and diffusion of dust, $C_{\rm surface}$ decreases exponentially with increasing ${\rm St}(z_{\rm s})/\alpha_z$, where 
$\alpha_z$ is the dimensionless parameter characterizing vertical diffusion strength (see Eq.~(A3) of \citealt{Okuzumi2025}). If ${\rm St}(z_{\rm s})/\alpha_z \gg 1$, the surface accretion layer is heavily depleted of dust $(C_{\rm surface} \approx 0)$, and one may approximate $\langle v_{{\rm d},r} \rangle_{\rm d,Sur}$ as
\begin{equation}
    \langle v_{{\rm d},r} \rangle_{\rm d,Sur}\approx \frac{\langle v^{\rm visc}_{{\rm g},r}\rangle _{\rm g}}{1+\rm St_{\rm mid}^2}  + \frac{2{\rm St_{\rm mid}}\Delta v_{\rm g,\phi,mid}}{1+\rm St_{\rm mid}^2}.
\label{eq:vdr_sur_appendix}
\end{equation}
For $z_{\rm s} = 2H_{\rm g}$ and $4H_{\rm g}$, where $H_{\rm g}=c_s/\Omega$ is the gas scale height, $C_{\rm surface}$ is below 0.1 when ${\rm St}_{\rm mid}/\alpha_z > 0.3$ and $10^{-3}$, respectively (see Appendix 1 and Fig.~8 of \citealt{Okuzumi2025}).

We adopt isotropic turbulence with $\alpha_r= \alpha_z=\alpha_{\rm visc}/3$ and $\alpha_{\rm visc}=3\times10^{-4}$, resulting in $\alpha_r=\alpha_z=1\times10^{-4}$.

\subsection{Sublimation }\label{sec:appendix_sublimation}
We simulate the evolution of total H$_2$O surface density $\Sigma_{\rm H_2O}=\Sigma_{\rm H_2O,g}+\Sigma_{\rm H_2O,d}$, where $\Sigma_{\rm H_2O,g}$ and $\Sigma_{\rm H_2O,d}$ are surface densities of H$_2$O vapors and ices.
We assume that sublimation and condensation occur instantaneously.
Then, the surface densities of vapor and ice phases are given by
\[
\Sigma_{\rm H_2O,g}(r) = 
\begin{cases}
  0 & (r \geq r_{\rm snowline})\\
  \Sigma_{\rm H_2O}(r)  &  (r < r_{\rm snowline}),
\end{cases}
\]
\[
\Sigma_{\rm H_2O,d}(r) = 
\begin{cases}
  \Sigma_{\rm H_2O}(r) & (r \geq r_{\rm snowline})\\
  0  &  (r < r_{\rm snowline}),
\end{cases}
\]
where $r_{\rm snowline}$ is fixed at the orbit where the disk temperature equals $T = 160$ K, corresponding to the sublimation temperature of water ice. 
Inside this radius, water exists primarily as vapor, while outside it remains as ice.
Using mass fluxes and source terms, the time evolution equations for the surface densities of ice and water vapor are given by
\begin{align*}
    \frac{\partial\Sigma_{\rm H_2O,d}}{\partial t}+\frac{1}{r}\frac{\partial}{\partial r}(rF_{\rm H_2O,d})&=S_{-}\\
    \frac{\partial\Sigma_{\rm H_2O,g}}{\partial t}+\frac{1}{r}\frac{\partial}{\partial r}(rF_{\rm H_2O,g})&=S_{+}.
\end{align*}
Here, $F$ is the mass flux of each phase, and $S_-$ and $S_+$ are the net source term representing the sublimation and condensation of $\rm H_2O$.
These source terms satisfy $S_- + S_+ = 0$ to ensure that the total mass of water is conserved during sublimation and condensation.
The time evolution equation for the surface density of total $\rm H_2O$ is given by
\begin{equation}\label{eq:Sigma_H2O_1}
    \frac{\partial\Sigma_{\rm H_2O}}{\partial t}+\frac{1}{r}\frac{\partial}{\partial r}\bigg[r(F_{\rm H_2O,d}+F_{\rm H_2O,g})\bigg]=0.
\end{equation}
We assume that vapor and ice originating from sublimation and condensation mix well with disk gas and dust, respectively.
This assumption allows us to adopt the radial velocity and the diffusion coefficient of gas and dust for H$_2$O vapors and ices.
In this case, $F_{\rm H_2O,d}$ and $F_{\rm H_2O,g}$ are given by
\begin{align*}
    F_{\rm H_2O,d}=\Sigma_{\rm H_2O,d}\langle v_{{\rm d},r} \rangle_{\rm d}- \Sigma_{\rm g} D_{{\rm d},r}\frac{\partial}{\partial r} \bigg(\frac{\Sigma_{\rm H_2O,d}}{\Sigma_{\rm g}}\bigg),\\
    F_{\rm H_2O,g}=\Sigma_{\rm H_2O,g}\langle v_{{\rm g},r} \rangle_{\rm g}- \Sigma_{\rm g} D_{{\rm g},r}\frac{\partial}{\partial r} \bigg(\frac{\Sigma_{\rm H_2O,g}}{\Sigma_{\rm g}}\bigg).
\end{align*}
Here, $\langle v_{{\rm g},r} \rangle_{\rm g}=\langle v^{\rm visc}_{{\rm g},r}\rangle_{\rm g} +\langle v^{\rm wind}_{{\rm g},r}\rangle_{\rm g}$.
In addition, assuming that ice particles are small and well coupled to the gas, the radial diffusion coefficient of $\mathrm{H_2O}$ can be regarded as nearly identical for each phase, i.e., $D_{{\rm d},r}\approx D_{{\rm g},r}=D_{{\rm H_2O},r}$.
Inserting the above flux formulation into Eq. \eqref{eq:Sigma_H2O_1}, the closed form of the evolution equation for $\mathrm{H_2O}$ surface density is given by
\begin{equation}
    \frac{\partial \Sigma_{\rm H_2O}}{\partial t} = -\frac{1}{r}\frac{\partial}{\partial r} r \bigg[\langle v_{{\rm H_2O},r} \rangle \Sigma_{\rm H_2O} - \Sigma_{\rm g} D_{{\rm H_2O},r}\frac{\partial}{\partial r} \bigg(\frac{\Sigma_{\rm H_2O}}{\Sigma_{\rm g}}\bigg ) \bigg ],
    \label{eq:h2o_surfacedensity}
\end{equation}
where we define the effective radial velocity of $\mathrm{H_2O}$ as
\begin{equation}
    \langle v_{{\rm H_2O},r} \rangle\equiv\frac{\Sigma_{\rm H_2O,d}\langle v_{{\rm d},r} \rangle_{\rm d}+\Sigma_{\rm H_2O,g}\langle v_{{\rm g},r} \rangle_{\rm g}}{\Sigma_{\rm H_2O}}.
\end{equation}
In practice, the radial velocity and diffusion coefficient of $\Sigma_{\rm H_2O}$ are equivalent to those of either dust or gas, depending on the orbital distance as expressed by
\[
\langle v_{{\rm H_2O},r} \rangle
 = 
\begin{cases}
  \langle v_{{\rm d},r} \rangle_{\rm d,(Uni\,or\,Sur)}, & r \geq r_{\rm snowline},\\
  \langle v_{{\rm g},r}, \rangle_{\rm g} &  r < r_{\rm snowline},
\end{cases}
\]
\[
D_{{\rm H_2O},r}
 = 
\begin{cases}
  D_{{\rm d},r}, & r \geq r_{\rm snowline},\\
  D_{{\rm g},r}, &  r < r_{\rm snowline}.
\end{cases}.
\]

\begin{table}[t]
 \centering
\begin{threeparttable}
  \caption{Summary of default model parameters}\label{table:1}
  \begin{tabular}{l r r r r} \hline
   Symbol & Description & Values \\ \hline \hline
    $M_{*}$& Stellar mass & $1M_{\odot}$ \\
    $M_{\rm disk,0}$ & Initial disk mass & $0.1M_{\odot}$\\
    $r_{c,0}$& Initial  characteristic disk radius & $30\,\rm au$\\
    $\alpha_{\rm visc}$ & Viscosity parameter & $3\,\times10^{-4}$\\
    $\alpha_{\rm wind}$ & Wind stress parameter & $6\,\times10^{-3}$\\
    $\lambda$ & Lever arm parameter & $3$\\
    $v_{\rm stick}$& Grain sticking threshold velocity &$\{0.3,1\}~\rm m~s^{-1}$\\
    $f_{\rm d}$ & Initial dust-to-gas mass ratio & $0.01$\\
    $f_{\rm H_2O}$ & Initial mass ratio of $\rm H_2O$ to dust & $0.5$\\
    \hline
  \end{tabular}
  \begin{tablenotes}
    \footnotesize
    \item Note:
    The same set of parameters is used for both the uniform-accretion and surface-accretion disk models.
    In our basic model, we adopt $v_{\rm stick} =0.3~\rm m~s^{-1}$.
  \end{tablenotes}

 \end{threeparttable} 

\end{table}

\subsection{Dust collision}
We also simulate the evolution of the column number density $N_{\rm d}$ taking into account collisional growth and fragmentation.
The column number density is used to compute the dust grains' mass $m_{\rm d} = (\Sigma_{\rm d} + \Sigma_{\rm H_2O,d})/N_{\rm d}$, radius $a = (3m_{\rm d}/(4\pi\rho_{\rm int}))^{1/3}$, and Stokes number ${\rm St}_{\rm mid}$ (Eq.~\eqref{eq:St}).

Following Okuzumi (2025), we solve the following equation to describe the evolution of $N_{\rm d}$:
\begin{equation}
    \frac{\partial N_{\rm d}}{\partial t} = -\frac{1}{r}\frac{\partial}{\partial r} r \bigg[\langle v_{{\rm d},r} \rangle_{\rm d} N_{\rm d} - \Sigma_{\rm g} D_{{\rm d},r}\frac{\partial}{\partial r} \bigg(\frac{N_{\rm d}}{\Sigma_{\rm g}}\bigg ) \bigg ]-\xi_{\rm stick}\frac{N_{\rm d}}{t_{\rm coll}}
    \label{eq:appendix_number_density},
\end{equation}
where $t_{\rm coll}$ is the mean collision time of the grains and $\xi_{\rm stick}$ is a dimensionless coefficient introduced to account for grain fragmentation at high collision speeds.

The mean collision time $t_{\rm coll}$ depends on the vertical distribution of the
dust density, which is characterized by the dust scale height $H_{\rm d} = (1+{\rm St_{mid}}/{\alpha_z})^{-0.5}H_{\rm g}$. 
Using $H_{\rm d}$ as described in \citet{Sato16}, $t_{\rm coll}$ is approximately given by
\begin{equation}
    t_{\rm coll} \approx \frac{H_{\rm d}}{2\sqrt{\pi}a^2\Delta v N_{\rm d}},
\end{equation}
where $a$ is the dust radius and $\Delta v$ is the dust collision velocity averaged over the grain vertical distribution. 
The collision velocity accounts for Brownian motion, radial and azimuthal drift, vertical settling, and turbulence \citep{Sato16}. 
As with turbulence diffusion, the turbulence-induced collision velocity is assumed to depend on $\alpha_{\rm visc}$ (see Sect.~4.2 of \citealt{Okuzumi2025}). Turbulence-induced collisional fragmentation limits ${\rm St}_{\rm mid}$ to $\approx v_{\rm stick}^2/(2.3\alpha_{\rm visc}c_{\rm s}^2)$ \citep{Birnstiel2009,OkuzumiTazaki19}}.

Following the formulation of \cite{Okuzumi2025}, we express $\xi_{\rm stick}$ as
\begin{equation}
    \xi_{\rm stick}= \min \left\{1,-\frac{\ln(\Delta v/v_{\rm stick})}{\ln 5}\right\},
\label{eq:appendix_xi}
\end{equation}
where $v_{\rm stick}$ is sticking threshold velocity below which colliding grains can grow in mass.

\begin{figure*}[t]
\centering
\includegraphics[
width=0.45\hsize,bb=0 0 580.4 382.5985]{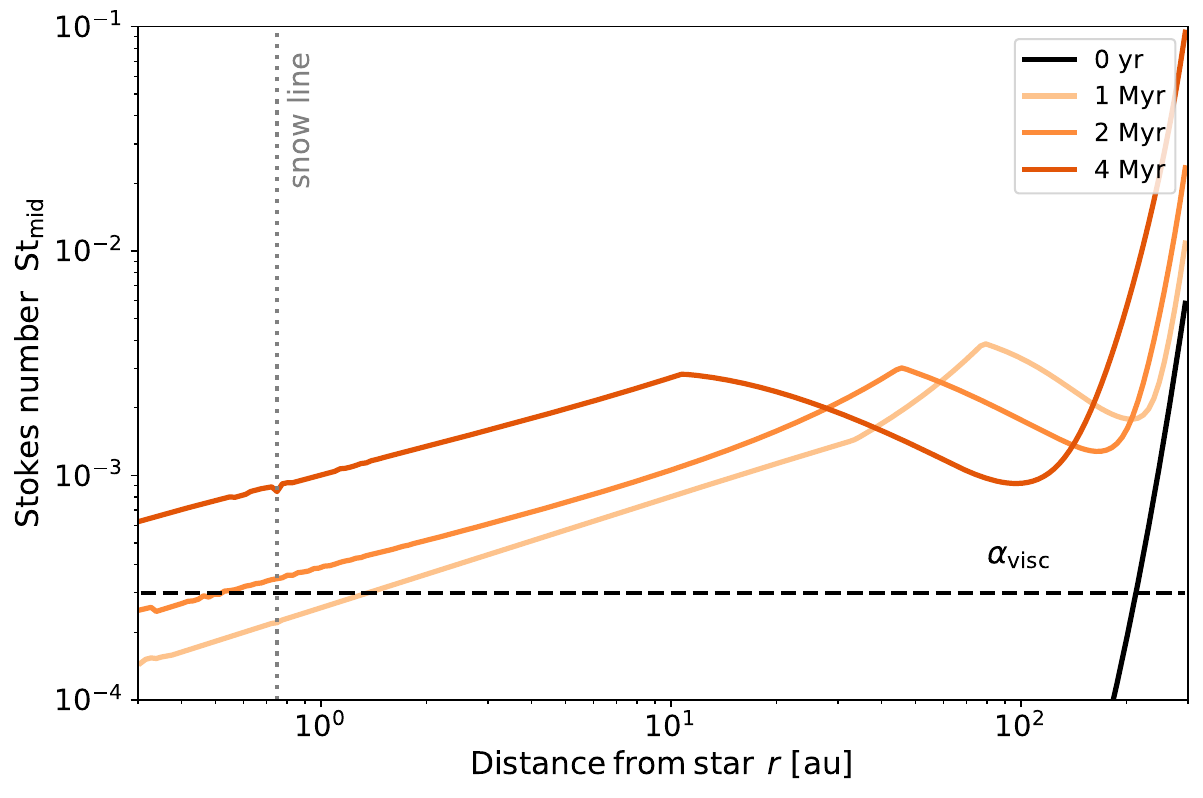}
\hspace{1mm}
\includegraphics[width=0.45\hsize,bb=0 0 580.4 382.5985]{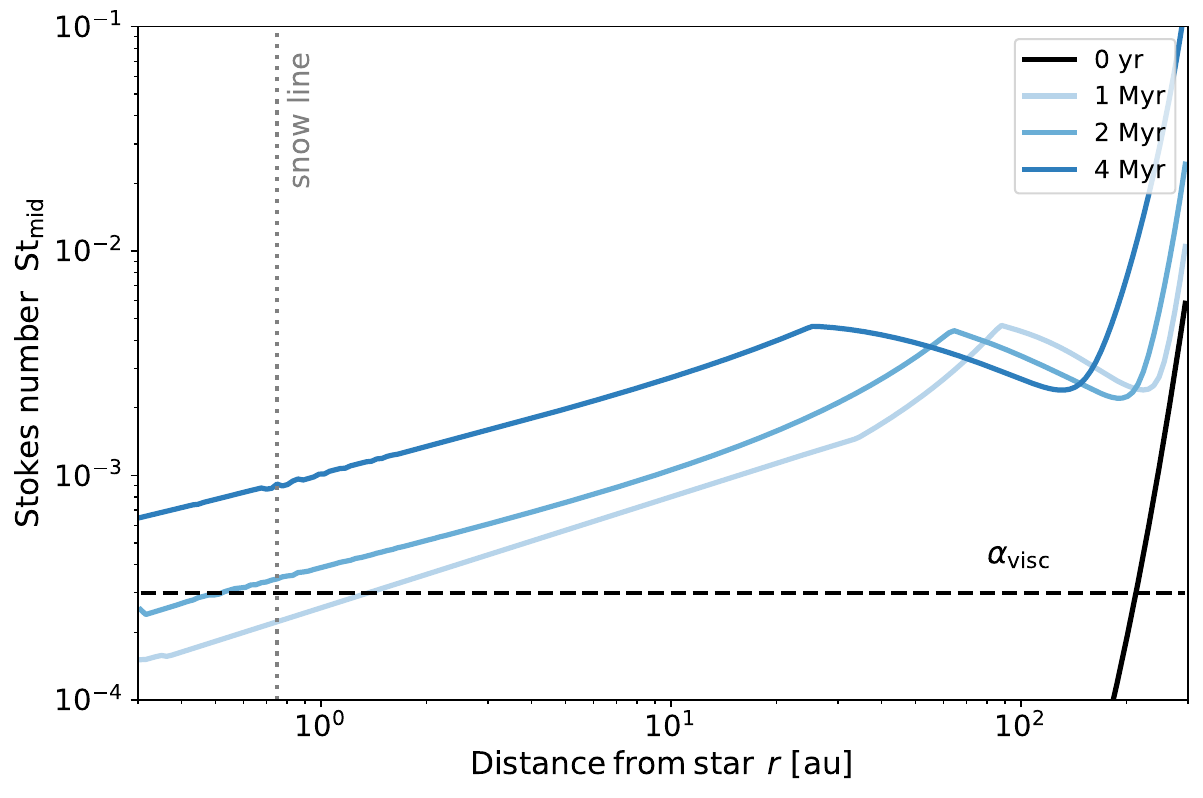}
\includegraphics[width=0.45\hsize,bb=0 0 580.4 382.5985]{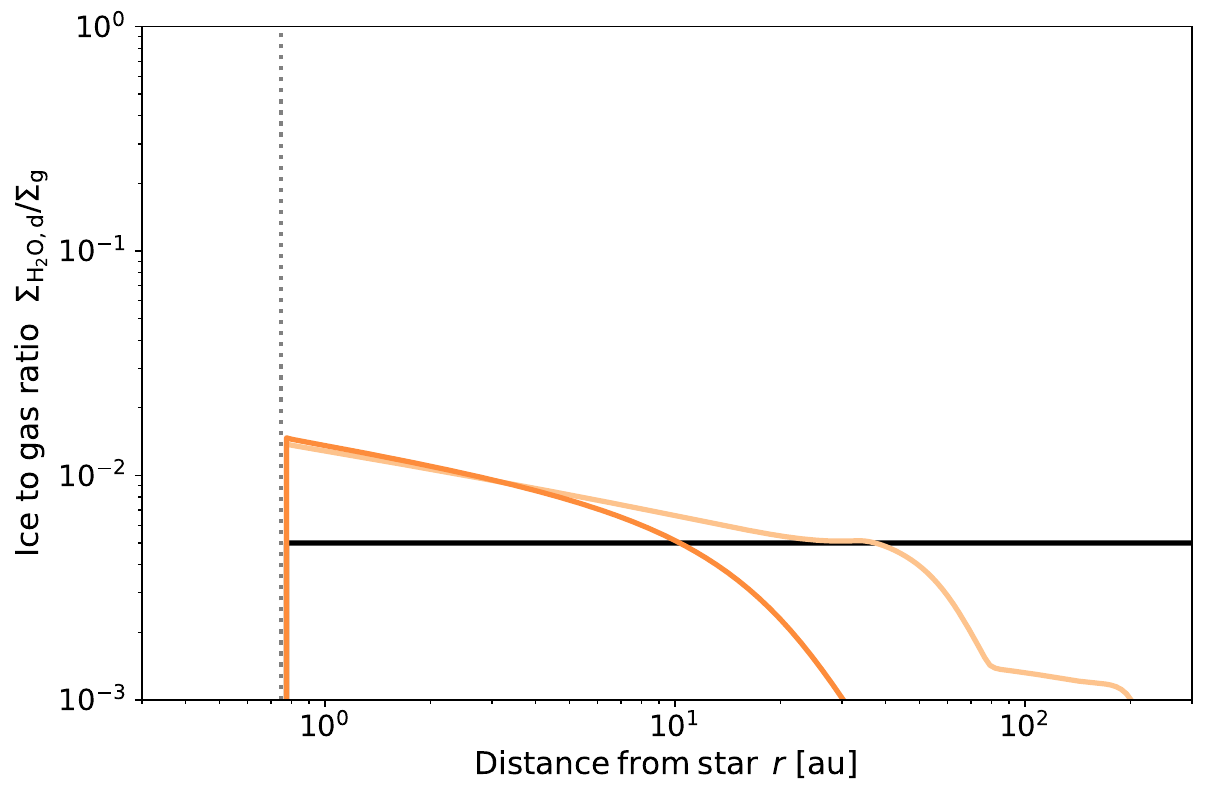}
\hspace{1mm}
\includegraphics[width=0.45\hsize,bb=0 0 580.4 382.5985]{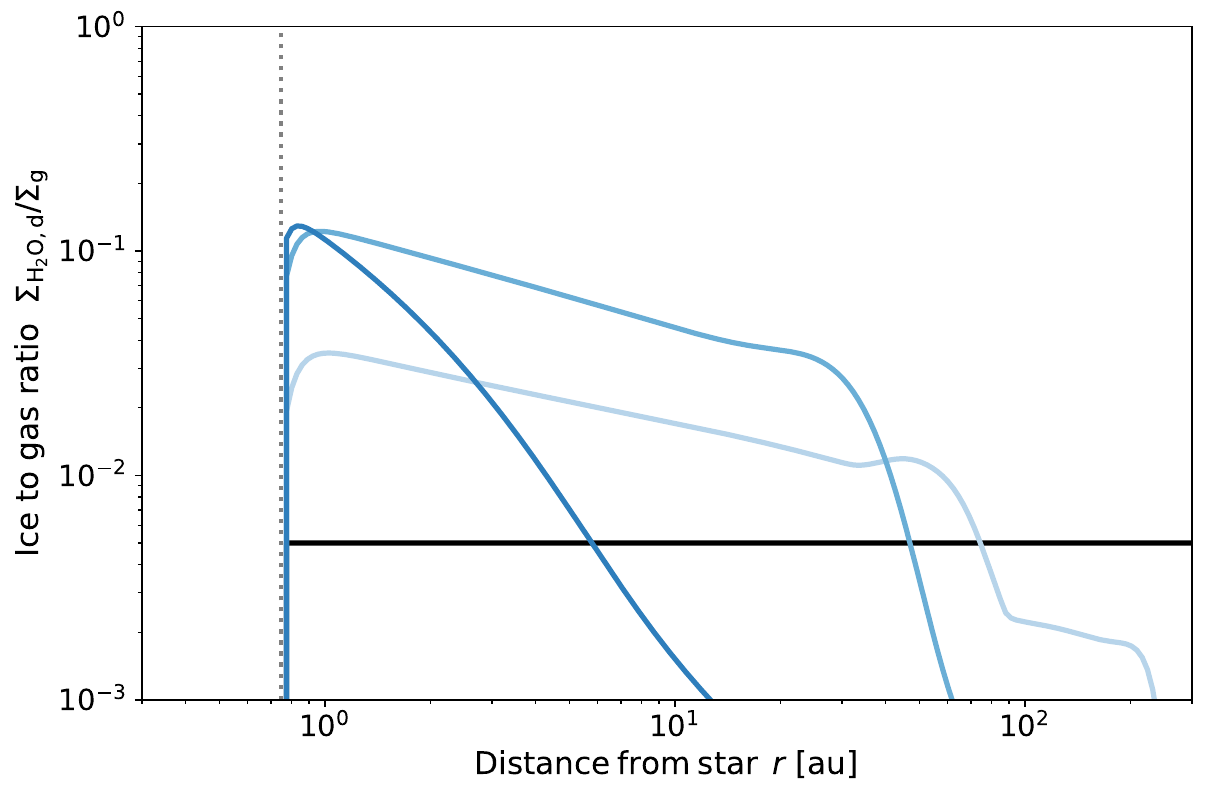}
\includegraphics[width=0.45\hsize,bb=0 0 580.4 382.5985]{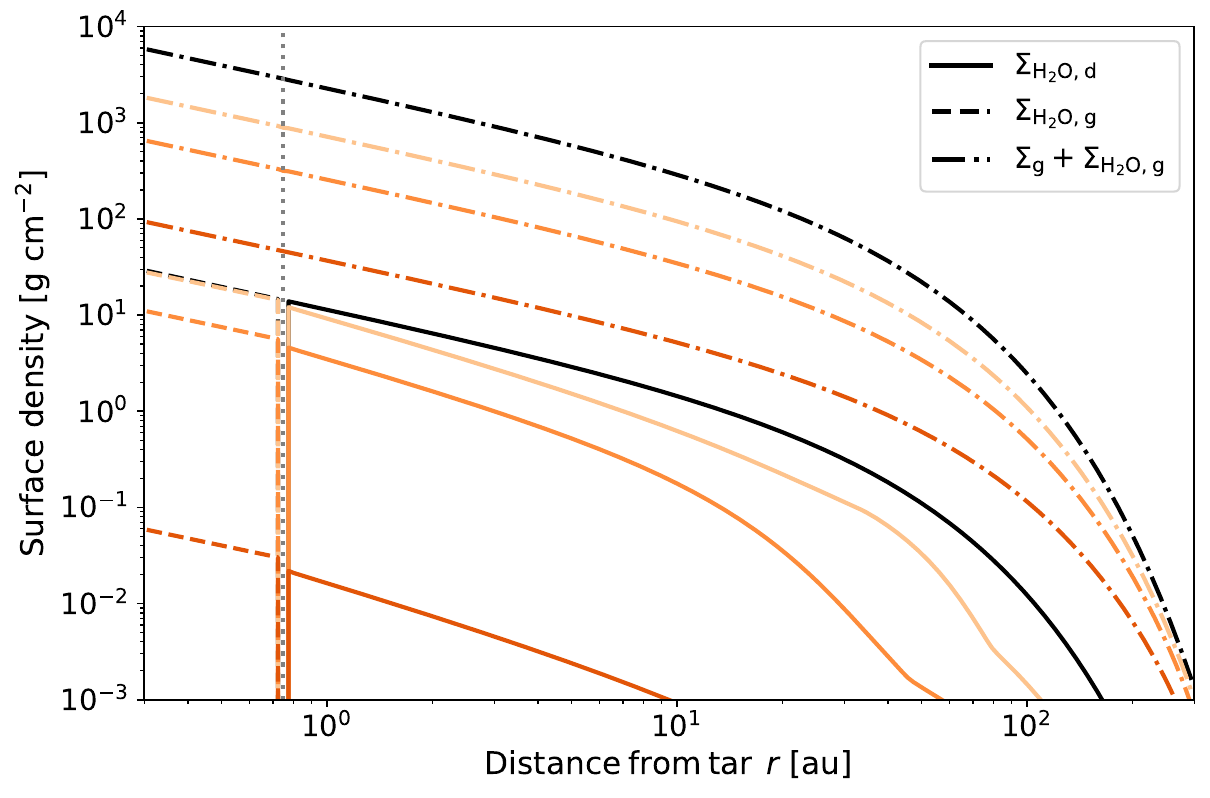}
\hspace{1mm}
\includegraphics[width=0.45\hsize,bb=0 0 580.4 382.5985]{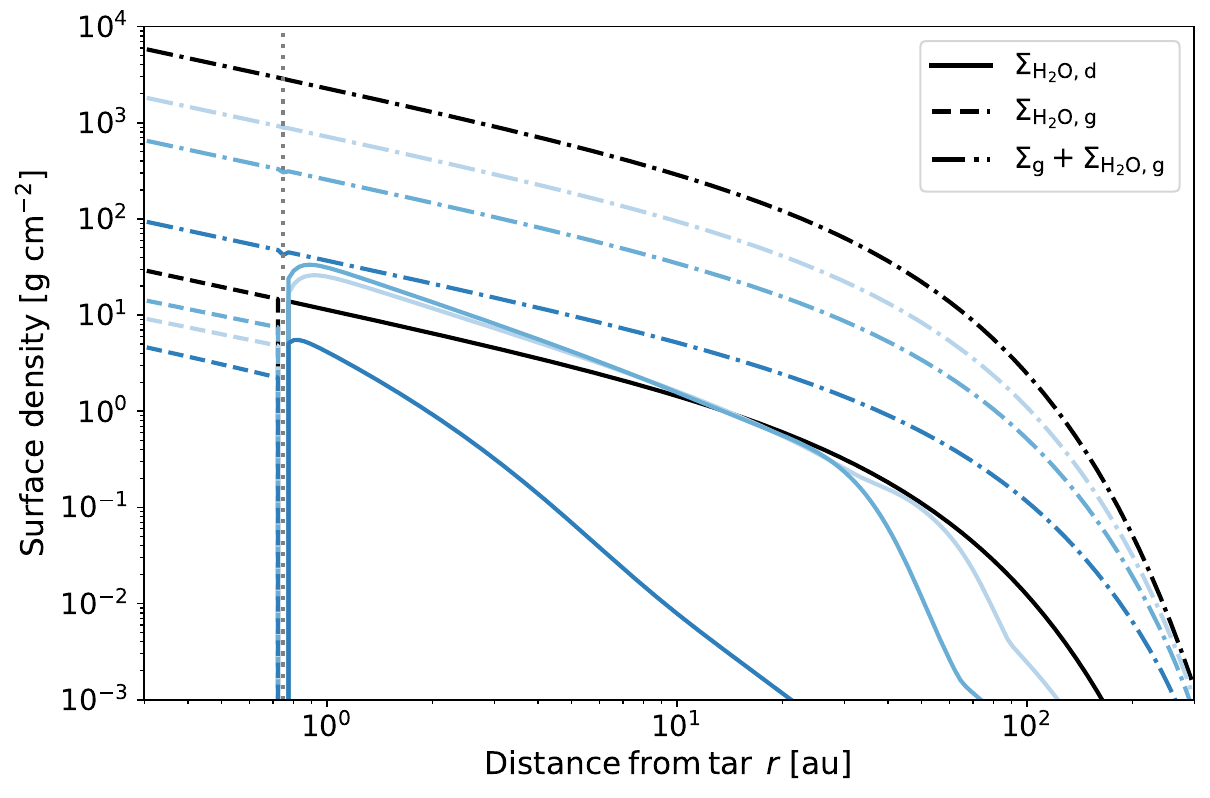}

\caption{
Evolution of disk radial profiles in uniform (left column) and surface (right column) accretion disks.
The top panels depict the evolution of the Stokes number St$_{\rm mid}$ of ice dust.
The middle panels show the evolution of the the ice-to-gas surface density ratio. The bottom panels show the evolution of the surface densities of ice, water vapor, and gas..
In the top panels, the black dashed line indicates the threshold above which drift dominates the transport of ice dust. Darker colors correspond to later times.
The initial profile is shown in black.
}
\label{fig:surface_density}
\end{figure*}

\subsection{Initial conditions}\label{sec:Numerical_model}
We numerically solve Eqs.~\eqref{eq:gas_surfacedensity_v}, \eqref{eq:dust_surfacedensity}, \eqref{eq:h2o_surfacedensity}, and \eqref{eq:appendix_number_density} which describe the evolution of the gas, dust, water components, and column number density, respectively.
The initial conditions are set to
\begin{equation*}
    \Sigma_{\rm g,0}(r) = \frac{M_{\rm disk,0}}{2\pi r^2_{\rm c,0}\Gamma(1+\xi)} \bigg( \frac{r}{r_{\rm c,0}}\bigg )^{-1+\xi}\exp \bigg(- \frac{r}{r_{\rm c,0}}\bigg ),
    \label{eq:initial_Sigma_g}
\end{equation*}
\begin{equation*}
    \Sigma_{\rm d,total,0} = f_{\rm d} \Sigma_{\rm g,0},
\end{equation*}
\begin{equation*}
    \Sigma_{\rm d,0} = (1-f_{\rm H_2O}) \Sigma_{\rm d,total,0},
\end{equation*}
\begin{equation*}
    \Sigma _{\rm H_2O,0} = f_{\rm H_2O} \Sigma_{\rm d,total,0},
\end{equation*}
where $f_{\rm d}$ denotes the initial dust-to-gas surface density ratio, while $f_{\rm H_2O}$ represents the initial surface density ratio of $\rm H_2O$ to dust.
In addition, $\Gamma$ denotes the Gamma function, $M_{\rm disk,0}$ is the initial disk gas mass, $M_\odot$ is the solar mass, $r_{c,0}$ is the initial characteristic radius of the disk gas, and $\xi = [2(\lambda-1)]^{-1}\alpha_{\rm wind}/(\alpha_{\rm wind}+\alpha_{\rm visc})$.
Table \ref{table:1} summarizes the adopted parameter values.
We assume an initial particle size of $a_0 = 10^{-5}~\rm cm$ throughout the disk. 
The initial Stokes number is determined from the Eq.~\eqref{eq:St}.
At the initial time $(t=0)$, the water vapor abundance in the disk gas is $\Sigma_{\rm H_2O,g,0}/\Sigma_{\rm g,0} = 0.5~\rm wt\% $.

\subsection{Simulated gas and dust evolution}\label{appendix:result}

Fig. \ref{fig:surface_density} shows the evolution of the dust Stokes number, the ice-to-gas mass ratio, and the surface densities of total gas, H$_2$O vapor, and ice in the uniform and surface-accretion models with $v_{\rm stick}=0.3~{\rm m~s^{-1}}$. 
At $r \ll 100~\rm au$, grains reach the fragmentation barrier, where ${\rm St}_{\rm mid} \approx v_{\rm stick}^2/(2.3\alpha_{\rm visc}c_{\rm s}^2)$.
The upper right panel of Fig.~\ref{fig:surface_density} confirms that the surface-accretion model satisfies the sufficient condition ${\rm St}_{\rm mid} > 0.3\alpha_z \approx 3\times 10^{-5}$ for dust settling below the accretion layer located at $z_{\rm s}\ge2H_{\rm g}$, which validates our assumption of $C_{\rm surface}\approx0$ (see Appendix~\ref{Appendix:gasdust}).
Note that previous MHD simulations showed the accretion layer typically located at $z_{\rm s}{\sim}2$--$4H_{\rm g}$ \citep[][]{Gressel15,Bai17,Lesur21,Iwasaki24}.
For $v_{\rm stick}=0.3~\rm m~s^{-1}$, this condition continues to hold as long as $\alpha_{\rm visc} \lesssim 3\times 10^{-3}$, which follows from the scaling ${\rm St}_{\rm mid}/\alpha_{z} \propto v_{\rm stick}^2/(\alpha_{z}\alpha_{\rm visc}) \propto v_{\rm stick}^2/\alpha_{\rm visc}^2$.

\section{Analytical estimates for water vapor enrichment}\label{Appendix:abundance}
Here, we provide a semi-analytical theory to interpret the water vapor enrichment in our basic model.
The water vapor sublimated at the snow line quickly mixes with the disk gas and flows inward.
Therefore, to understand the water vapor enrichment in the disk gas, one can look at the ice-to-gas mass accretion ratio just outside the snow line.
Given the same radial velocity for H$_2$O vapor and the rest of the disk gas, in a steady state, mass conservation demands the following relation between the H$_2$O vapor enrichment inside the snow line and the ice-to-gas mass accretion ratio outside the snow line:
\begin{equation}\label{eq:Appendix1}
    \frac{\Sigma_{\rm H_2O,g}}{\Sigma_{\rm g}}\Bigg|_{r<r_{\rm snow line}}=\frac{\dot{M}_{\rm H_2O,d}}{\dot{M}_{\rm g}}\Bigg|_{r>r_{\rm snowline}}=\frac{\Sigma_{\rm H_2O,d}}{\Sigma_{\rm g}}\frac{\langle v_{{\rm H_2O},r} \rangle}{\langle v_{{\rm g},r}\rangle_{\rm g}}.
\end{equation}
Here, $\dot{M}_{\rm H_2O,d}$ and $\dot{M}_{\rm g}$ are the mass accretion rates of water ice and gas, respectively.
For convenience, we rewrite Eq. \eqref{eq:Appendix1} in terms of $\alpha_{\rm visc}$, $\alpha_{\rm wind}$, and $\rm St_{\rm mid}$.
We assume that the disk evolves quasi-steadily near the snow line and adopt a geometrically thin, vertically isothermal Keplerian disk. The sound speed scales as $c_s \propto \sqrt{T}\propto r^{-1/4}$, while the gas surface density follows $\Sigma_{\rm g}\propto r^{-1+\xi}$ at $r\ll r_{\rm c,0}$.
For our basic parameters, we obtain $\xi\approx1/4$.
Using these values, Eqs.~\eqref{eq:vgr_visc} and \eqref{eq:vgr_wind}, as well as $\Delta v_{\rm g,\phi,mid}$, can be expressed in a scaled form.
Therefore, the following approximations can be made for each case:
\begin{align}
     \langle v^{\rm visc}_{{\rm g},r}\rangle_{\rm g}\approx-\frac{9}{4}\alpha_{\rm visc}\Big(\frac{H_{\rm g}}{r}\Big)^{2}r\Omega,\\
     \langle v^{\rm wind}_{{\rm g},r}\rangle_{\rm g}\approx-\frac{3}{2}\alpha_{\rm wind}\Big(\frac{H_{\rm g}}{r}\Big)^{2}r\Omega,\\
     \Delta v_{\rm g,\phi,mid}\approx-\frac{5}{4}\Big(\frac{H_{\rm g}}{r}\Big)^{2}r\Omega.
     \label{eq:delta_v}
\end{align}
Inserting the above equations and Eq.~\eqref{eq:vdr_uni_appendix} with $\rm St_{\rm mid} \ll 1$ into Eq. \eqref{eq:Appendix1}, the mass accretion rate ratio in a uniform accretion disk can be expressed as
\begin{equation}
     \frac{\dot{M}_{\rm H_2O,d}}{\dot{M}_{\rm g}}\Bigg|_{\rm Uni}\approx\frac{\Sigma_{\rm H_2O,d}}{\Sigma_{\rm g}}\frac{(\frac{9}{10}\alpha_{\rm visc}+\frac{3}{5}\alpha_{\rm wind} )+\rm St_{\rm mid}}{\frac{9}{10}\alpha_{\rm visc}+\frac{3}{5}\alpha_{\rm wind}}.
\end{equation}
The first term in the velocity ratio represents the gas dragging the ice dust, while the second term corresponds to the drift of the ice dust itself.
As shown in Table~\ref{table:1}, we fix $\alpha_{\rm visc}$ and $\alpha_{\rm wind}$, with their ratio set to $\alpha_{\rm wind}/\alpha_{\rm visc} = 20$.
Moreover, as seen in the top-left panel of Fig.~\ref{fig:surface_density}, $\rm St_{\rm mid}$ near the snow line never exceeds the order of $(9/10)\alpha_{\rm visc} + (3/5)\alpha_{\rm wind}{\sim}\alpha_{\rm wind}=6\times10^{-3}$ before the disk reaches the end of its lifetime.
Therefore, under the assumption of fragile dust in a uniform accretion disk, the $\rm H_2O$ vapor enrichment inside the snow line can be well approximated as
\begin{equation}
    \frac{\Sigma_{\rm H_2O,g}}{\Sigma_{\rm g}}\Bigg|_{\rm Uni}=\frac{\dot{M}_{\rm H_2O,d}}{\dot{M}_{\rm g}}\Bigg|_{\rm Uni}\approx\frac{\Sigma_{\rm H_2O,d}}{\Sigma_{\rm g}}\label{eq:enrich_uniform}.
\end{equation}
Thus, the vapor enrichment in the uniform accreting disk approximately follows the ice-to-gas mass ratio right outside the H$_2$O snow line.

In the surface accreting disk, since dust transport through wind-driven gas flow is negligible, the mass accretion rate ratio can be expressed as 
\begin{equation}
     \frac{\dot{M}_{\rm H_2O,d}}{\dot{M}_{\rm g}}\Bigg|_{\rm Sur}\approx\frac{\Sigma_{\rm H_2O,d}}{\Sigma_{\rm g}}\frac{\frac{9}{10}\alpha_{\rm visc} +\rm St_{\rm mid}}{\frac{9}{10}\alpha_{\rm visc}+\frac{3}{5}\alpha_{\rm wind}}.
\end{equation}
Thus, inserting our fiducial value of $\alpha_{\rm wind}/\alpha_{\rm visc} = 20$, the final mass accretion rate ratio in a surface accreting disk can be expressed as
\begin{equation}\label{eq:enrich_surface}
    \frac{\Sigma_{\rm H_2O,g}}{\Sigma_{\rm g}}\Bigg|_{\rm Sur}=\left.\frac{\dot{M}_{\rm H_2O,d}}{\dot{M}_{\rm g}}\right|_{\rm Sur}\approx \frac{\Sigma_{\rm H_2O,d}}{\Sigma_{\rm g}} \frac{10}{129}\left(\frac{9}{10}+\frac{\rm St_{mid}}{\alpha_{\rm visc}}\right).
\end{equation}
In our simulation, $\rm St_{mid}$ remains in the range of $0.5<\rm St_{mid}/\alpha_{visc}<3$ near the snow line during $0$--$4$~Myr (the top-right panel of Fig.~\ref{fig:surface_density}).
Therefore, the vapor enrichment inside the snow line takes a value of $\sim0.1$--$0.3\times \Sigma_{\rm H_2O,d}/\Sigma_{\rm g}$ according to Eq.~\eqref{eq:enrich_surface}.

In contrast, the ice-to-gas surface density ratio changes over a range of approximately $0.01$--$0.1$ (the middle-right panel of Fig.~\ref{fig:surface_density}).
While ${\rm St}_{\rm mid}/\alpha_{\rm visc}$ undergoes a factor of $\sim3$ enhancement over 4 Myr, the ice-to-gas surface density ratio experiences more than an order of magnitude enhancement.
Therefore, in all models, for fragile dust, the qualitative evolution of water vapor enrichment is largely determined by the ice-to-gas surface density ratio (the middle-panels of Fig.~\ref{fig:surface_density})

In our models, the evolution of the gas surface density is nearly identical in both cases (the bottom panels of Fig.~\ref{fig:surface_density}).
Therefore, Eqs.~\eqref{eq:enrich_uniform} and \eqref{eq:enrich_surface} indicate that the difference in the water vapor enrichment originates primarily from differences in the evolution of $\Sigma_{\rm H_2O,d}$ (the middle and bottom-panels of Fig.~\ref{fig:surface_density}).
In a uniform accretion disk, since the ice dust retains ${\rm St}_{\rm mid}\lesssim \alpha_{\rm visc}+\alpha_{\rm wind}$ throughout 4 Myr, the surface density of the ice dust co-evolves with the surface density of the gas, as illustrated in the middle and bottom-left panel of Fig.~\ref{fig:surface_density}.
This explains why the ice-to-gas surface density ratio remains roughly constant at $t\lesssim 2$ Myr, causing an approximately constant water vapor concentration until the exhaustion of the ice-bearing dust (see the top panel of Fig.

In contrast, in a surface accreting disk where wind-driven gas flow does not affect dust transport, ice-bearing dust falls onto the central star more slowly than the gas if dust is fragile and thus small.
As a result, the ice-to-gas surface density ratio $\Sigma_{\rm H_2O,d}/\Sigma_{\rm g}$ increases with time, as seen in the middle right panel of Fig. \ref{fig:surface_density}.
The accumulation of ice-bearing dust progressively causes a water vapor enrichment as the disk gas mass decreases, resulting in the anti-correlation between the water vapor enrichment and the disk gas mass, as illustrated in the upper panel of Fig.~\ref{fig:enrichment_to_initial}.


\begin{thebibliography}{32}
\expandafter\ifx\csname natexlab\endcsname\relax\def\natexlab#1{#1}\fi

\bibitem[{{Arulanantham} {et~al.}(2025){Arulanantham}, {Salyk}, {Pontoppidan},
  {Banzatti}, {Zhang}, {{\"O}berg}, {Long}, {Carr}, {Najita}, {Pascucci},
  {Colmenares}, {Xie}, {Huang}, {Green}, {Andrews}, {Blake}, {Bergin},
  {Pinilla}, {Vioque}, {Dahl}, {Raul}, {Krijt}, \& {The Jdiscs
  Collaboration}}]{Arulanantham2025}
{Arulanantham}, N., {Salyk}, C., {Pontoppidan}, K., {et~al.} 2025, \aj, 170, 67

\bibitem[{{Asplund} {et~al.}(2009){Asplund}, {Grevesse}, {Sauval}, \&
  {Scott}}]{Asplund2009}
{Asplund}, M., {Grevesse}, N., {Sauval}, A.~J., \& {Scott}, P. 2009, \araa, 47,
  481

\bibitem[{{Bai}(2017)}]{Bai17}
{Bai}, X.-N. 2017, \apj, 845, 75

\bibitem[{{Banzatti} {et~al.}(2025){Banzatti}, {Salyk}, {Pontoppidan}, {Carr},
  {Zhang}, {Arulanantham}, {Krijt}, {{\"O}berg}, {Cleeves}, {Najita},
  {Pascucci}, {Blake}, {Romero-Mirza}, {Bergin}, {Cieza}, {Pinilla}, {Long},
  {Mallaney}, {Xie}, {Waggoner}, {Kaeufer}, \& {The Jdiscs
  Collaboration}}]{Banzatti2025}
{Banzatti}, A., {Salyk}, C., {Pontoppidan}, K.~M., {et~al.} 2025, \aj, 169, 165

\bibitem[{{Bean} {et~al.}(2023){Bean}, {Xue}, {August}, {Lunine}, {Zhang},
  {Thorngren}, {Tsai}, {Stassun}, {Schlawin}, {Ahrer}, {Ih}, \&
  {Mansfield}}]{Bean2023}
{Bean}, J.~L., {Xue}, Q., {August}, P.~C., {et~al.} 2023, \nat, 618, 43

\bibitem[{{Birnstiel} {et~al.}(2009){Birnstiel}, {Dullemond}, \&
  {Brauer}}]{Birnstiel2009}
{Birnstiel}, T., {Dullemond}, C.~P., \& {Brauer}, F. 2009, \aap, 503, L5

\bibitem[{{Booth} {et~al.}(2017){Booth}, {Clarke}, {Madhusudhan}, \&
  {Ilee}}]{Booth2017}
{Booth}, R.~A., {Clarke}, C.~J., {Madhusudhan}, N., \& {Ilee}, J.~D. 2017,
  \mnras, 469, 3994

\bibitem[{{Dominik} \& {Tielens}(1997)}]{DominikTielens97}
{Dominik}, C. \& {Tielens}, A.~G.~G.~M. 1997, \apj, 480, 647

\bibitem[{{Feinstein} {et~al.}(2023){Feinstein}, {Radica}, {Welbanks},
  {Murray}, {Ohno}, {Coulombe}, {Espinoza}, {Bean}, {Teske}, {Benneke}, {Line},
  {Rustamkulov}, {Saba}, {Tsiaras}, {Barstow}, {Fortney}, {Gao}, {Knutson},
  {MacDonald}, {Mikal-Evans}, {Rackham}, {Taylor}, {Parmentier}, {Batalha},
  {Berta-Thompson}, {Carter}, {Changeat}, {dos Santos}, {Gibson}, {Goyal},
  {Kreidberg}, {L{\'o}pez-Morales}, {Lothringer}, {Miguel}, {Molaverdikhani},
  {Moran}, {Morello}, {Mukherjee}, {Sing}, {Stevenson}, {Wakeford}, {Ahrer},
  {Alam}, {Alderson}, {Allen}, {Batalha}, {Bell}, {Blecic}, {Brande},
  {Caceres}, {Casewell}, {Chubb}, {Crossfield}, {Crouzet}, {Cubillos}, {Decin},
  {D{\'e}sert}, {Harrington}, {Heng}, {Henning}, {Iro}, {Kempton}, {Kendrew},
  {Kirk}, {Krick}, {Lagage}, {Lendl}, {Mancini}, {Mansfield}, {May}, {Mayne},
  {Nikolov}, {Palle}, {Petit dit de la Roche}, {Piaulet}, {Powell}, {Redfield},
  {Rogers}, {Roman}, {Roy}, {Nixon}, {Schlawin}, {Tan}, {Tremblin}, {Turner},
  {Venot}, {Waalkes}, {Wheatley}, \& {Zhang}}]{Feinstein2023}
{Feinstein}, A.~D., {Radica}, M., {Welbanks}, L., {et~al.} 2023, \nat, 614, 670

\bibitem[{{Fortney} {et~al.}(2013){Fortney}, {Mordasini}, {Nettelmann},
  {Kempton}, {Greene}, \& {Zahnle}}]{Fortney2013}
{Fortney}, J.~J., {Mordasini}, C., {Nettelmann}, N., {et~al.} 2013, \apj, 775,
  80

\bibitem[{{Fu} {et~al.}(2025){Fu}, {Stevenson}, {Sing}, {Mukherjee},
  {Welbanks}, {Thorngren}, {Tsai}, {Gao}, {Lothringer}, {Beatty}, {Gapp},
  {Evans-Soma}, {Allart}, {Pelletier}, {Thao}, \& {Mann}}]{Fu2025}
{Fu}, G., {Stevenson}, K.~B., {Sing}, D.~K., {et~al.} 2025, \apj, 986, 1

\bibitem[{{Fu} {et~al.}(2024){Fu}, {Welbanks}, {Deming}, {Inglis}, {Zhang},
  {Lothringer}, {Ih}, {Moses}, {Schlawin}, {Knutson}, {Henry}, {Greene},
  {Sing}, {Savel}, {Kempton}, {Louie}, {Line}, \& {Nixon}}]{Fu2024}
{Fu}, G., {Welbanks}, L., {Deming}, D., {et~al.} 2024, \nat, 632, 752

\bibitem[{{Gressel} {et~al.}(2015){Gressel}, {Turner}, {Nelson}, \&
  {McNally}}]{Gressel15}
{Gressel}, O., {Turner}, N.~J., {Nelson}, R.~P., \& {McNally}, C.~P. 2015,
  \apj, 801, 84

\bibitem[{{Gundlach} {et~al.}(2018){Gundlach}, {Schmidt}, {Kreuzig},
  {Bischoff}, {Rezaei}, {Kothe}, {Blum}, {Grzesik}, \& {Stoll}}]{Gundlach2018}
{Gundlach}, B., {Schmidt}, K.~P., {Kreuzig}, C., {et~al.} 2018, \mnras, 479,
  1273

\bibitem[{{Hu} \& {Bai}(2021)}]{Fu&Bai2021}
{Hu}, Z. \& {Bai}, X.-N. 2021, \mnras, 503, 162

\bibitem[{{Iwasaki} {et~al.}(2024){Iwasaki}, {Tomida}, {Takasao}, {Okuzumi}, \&
  {Suzuki}}]{Iwasaki24}
{Iwasaki}, K., {Tomida}, K., {Takasao}, S., {Okuzumi}, S., \& {Suzuki}, T.~K.
  2024, \pasj, 76, 616

\bibitem[{{Jiang} {et~al.}(2024){Jiang}, {Mac{\'\i}as}, {Guerra-Alvarado}, \&
  {Carrasco-Gonz{\'a}lez}}]{Jiang24}
{Jiang}, H., {Mac{\'\i}as}, E., {Guerra-Alvarado}, O.~M., \&
  {Carrasco-Gonz{\'a}lez}, C. 2024, \aap, 682, A32

\bibitem[{{Kempton} \& {Knutson}(2024)}]{Kempton24}
{Kempton}, E. M.~R. \& {Knutson}, H.~A. 2024, Reviews in Mineralogy and
  Geochemistry, 90, 411

\bibitem[{{Lesur}(2021)}]{Lesur21}
{Lesur}, G. R.~J. 2021, \aap, 650, A35

\bibitem[{{Long} {et~al.}(2025){Long}, {Pascucci}, {Houge}, {Banzatti},
  {Pontoppidan}, {Najita}, {Krijt}, {Xie}, {Williams}, {Herczeg}, {Andrews},
  {Bergin}, {Blake}, {Colmenares}, {Harsono}, {Romero-Mirza}, {Li}, {Lu},
  {Pinilla}, {Wilner}, {Vioque}, {Zhang}, \& {JDISCS Collaboration}}]{Long2025}
{Long}, F., {Pascucci}, I., {Houge}, A., {et~al.} 2025, \apjl, 978, L30

\bibitem[{{Lothringer} {et~al.}(2026){Lothringer}, {Lowson}, \&
  {Fu}}]{Lothringer2026}
{Lothringer}, J.~D., {Lowson}, N., \& {Fu}, G. 2026, \aj, 171, 31

\bibitem[{{Musiolik} \& {Wurm}(2019)}]{Musiolik2019}
{Musiolik}, G. \& {Wurm}, G. 2019, \apj, 873, 58

\bibitem[{{Ohno} {et~al.}(2026){Ohno}, {Ikoma}, {Okuzumi}, \&
  {Kimura}}]{Ohno2026}
{Ohno}, K., {Ikoma}, M., {Okuzumi}, S., \& {Kimura}, T. 2026, \pasj, 78, 493

\bibitem[{{Okuzumi}(2025)}]{Okuzumi2025}
{Okuzumi}, S. 2025, \pasj, 77, 162

\bibitem[{{Okuzumi} \& {Tazaki}(2019)}]{OkuzumiTazaki19}
{Okuzumi}, S. \& {Tazaki}, R. 2019, \apj, 878, 132

\bibitem[{{Sato} {et~al.}(2016){Sato}, {Okuzumi}, \& {Ida}}]{Sato16}
{Sato}, T., {Okuzumi}, S., \& {Ida}, S. 2016, \aap, 589, A15

\bibitem[{{Schneider} \& {Bitsch}(2021)}]{Schneider21}
{Schneider}, A.~D. \& {Bitsch}, B. 2021, \aap, 654, A71

\bibitem[{{Shakura} \& {Sunyaev}(1973)}]{Shakura73}
{Shakura}, N.~I. \& {Sunyaev}, R.~A. 1973, \aap, 24, 337

\bibitem[{{Shibata} {et~al.}(2023){Shibata}, {Helled}, \&
  {Kobayashi}}]{Shibata2023}
{Shibata}, S., {Helled}, R., \& {Kobayashi}, H. 2023, \mnras, 519, 1713

\bibitem[{{Tabone} {et~al.}(2022){Tabone}, {Rosotti}, {Cridland}, {Armitage},
  \& {Lodato}}]{Tabone22}
{Tabone}, B., {Rosotti}, G.~P., {Cridland}, A.~J., {Armitage}, P.~J., \&
  {Lodato}, G. 2022, \mnras, 512, 2290

\bibitem[{{Ueda} {et~al.}(2024){Ueda}, {Tazaki}, {Okuzumi}, {Flock}, \&
  {Sudarshan}}]{Ueda24}
{Ueda}, T., {Tazaki}, R., {Okuzumi}, S., {Flock}, M., \& {Sudarshan}, P. 2024,
  Nature Astronomy, 8, 1148

\bibitem[{{Welbanks} {et~al.}(2019){Welbanks}, {Madhusudhan}, {Allard},
  {Hubeny}, {Spiegelman}, \& {Leininger}}]{Welbanks19}
{Welbanks}, L., {Madhusudhan}, N., {Allard}, N.~F., {et~al.} 2019, \apjl, 887,
  L20

\end{thebibliography}
\end{document}